
%
%
%
%
%
%
%

\magnification 1200
\baselineskip = 17pt

\noindent{\bf LA-UR-94-94}
\bigskip
\centerline{\bf DIRECT CALCULATION OF SPIN-STIFFNESS}
\bigskip
\centerline{\bf FOR SPIN-1/2 HEISENBERG MODELS}
\vskip 50pt
\centerline{J. Bon\v ca,\footnote{$^1$}{Permanent address:
J. Stefan Inst., Univ. of Ljubljana, 61111 Ljubljana, Slovenia.}
J. P. Rodriguez,\footnote{$^2$}{Permanent address:
Dept. of Physics and Astronomy,
California State University,
Los Angeles, CA 90032.}
J. Ferrer,\footnote{$^3$}{Permanent address:
Dpto. de Fisica de la Materia Condensada (C-12),
Universidad Autonoma de Madrid, Cantoblanco, Spain.}
K.S. Bedell}
\medskip
\centerline{Theoretical Division and CNLS,
Los Alamos National Laboratory, Los Alamos, NM 87545.}
\vskip 30pt
\centerline  {\bf  Abstract}
\vskip 8pt\noindent
The spin-stiffness of frustrated spin-1/2 Heisenberg models in one and
two dimensions is computed for the first time by exact
diagonalizations on small clusters that implement spin-dependent
twisted boundary conditions.  Finite-size extrapolation to the
thermodynamic limit yields a value of $0.14\pm 0.01$ for the
spin-stiffness of the unfrustrated planar antiferromagnet.  We also
present a general discussion of the linear-response theory for
spin-twists, which ultimately leads to the moment sum-rule.

\bigskip
\noindent
PACS Indices:  75.10.Jm, 75.50.Ee, 75.30.Cr, 74.20.Mn

\vfill\eject
One of the most basic questions in the study of magnetism is that of
the existence or absence of long-range order in the corresponding
magnetic moments.  A partial answer to this question can be given by
the determination of the so-called spin-stiffness of the magnet, which
measure the rigidity of the spins with respect to a small twist.  In
particular, systems possessing long-range spin-order are stiff, while
spin systems that are not stiff accordingly show no long-range order
in the moments.  In the case of spin-1/2 systems, the latter stiffness
can be directly measured by the generation of a spin-current with a
spin-dependent magnetic field, as was first shown by Shastry and
Sutherland.$^1$ This method is analogous to that used to measure the
charge-stiffness of a system,$^2$ which discriminates between metals
and insulators.

In this paper, we apply the former method to the case of the spin-1/2
Heisenberg model, $H_0 = \sum_{( {\bf i,j})}J_{\bf ij}\vec S_{\bf
i}\cdot\vec S_{\bf j}$, on both chain and square-lattice geometries.
First, we give a general discussion of spin-twists for this model
based on linear-response theory, which ultimately leads to the moment
sum-rule.$^3$ We then make practical use of the above method to
measure the spin-stiffness of near-neighbor Heisenberg chains and
square lattices by exact diagonalization of the $S_z=0$ subspace with
the Lanczos technique.$^4$ Employing finite-size extrapolations, we
find values of the spin-stiffness for the nearest-neighbor Heisenberg
ferromagnet and antiferromagnet on the square lattice that agree with
spin-wave theory results to within ten percent.$^5$ In the particular
case of the square-lattice, where reported results for this quantity
vary widely,$^6$ we obtain an upper bound for the spin-stiffness of
$\rho_s/J\cong 0.174$, as well as an extrapolated value of
$\rho_s/J=0.14\pm 0.01$ in the thermodynamic limit.  Also, for the
case of spin-1/2 frustrated antiferromagnets with next (next)
nearest-neighbor interactions, we generally find that the stiffness
coefficient vanishes near the point where the analogous classical
model losses long-range order in the magnetic moments.$^7$ We now turn
to the derivation of the moment sum-rule.

{\it Linear-Response.} Following ref. 1, the rigidity of a Heisenberg
model with respect to a twist about the spin z-axis is reflected in
the ground-state energy of the modified Hamiltonian,
$$H = \sum_{({\bf i,j})}J_{\bf ij}\Biggl[{1\over 2}
(S_{\bf i}^+S_{\bf j}^- e^{i\theta_{\bf ij}}
+S_{\bf i}^-S_{\bf j}^+ e^{-i\theta_{\bf ij}})
+S_{\bf i}^zS_{\bf j}^z\Biggr],
\eqno (1)
$$ where
$\theta_{\bf ij}$ represents the twist angle on the bond $({\bf
i,j})$.  Hence, in the limit of small twists, the above Hamiltonian is
expressible as $H = H_0+H_1$, where $H_0$ represents the unperturbed
Heisenberg model with $\theta_{\bf ij} = 0$, and where the perturbation to
this Hamiltonian is given by
$$H_1 = \sum_{({\bf i,j})}[\theta_{\bf ij} j_{\bf ij}^{(s)}
-{1\over 4}\theta_{\bf ij}^2J_{\bf ij}(S_{\bf i}^+S_{\bf j}^-
+ {\rm h.c.})].
\eqno (2)
$$
Above, $j_{\bf ij}^{(s)} = {i\over 2}J_{\bf ij}(S_{\bf i}^+S_{\bf j}^-
- {\rm h.c.})$ is the $z$-component of the spin-current operator.
Consider now the case of identical twists, $\theta_x$, that exist only
along nearest-neighbor bonds oriented along the $x$-axis.  Then since
the spin-rigidity, $D_s$, is related to the difference in the
ground-state energy by $E_0(\theta_x)-E_0(0) = N D_s\theta_x^2$ for
small $\theta_x$, second-order perturbation theory gives
$$D_s = N^{-1}\Biggl({1\over{2}}\langle -T_x^{(s)}\rangle -
\sum_{\nu\ne 0}{|\langle
0|j_x^{(s)}|\nu\rangle|^2\over{E_{\nu}-E_0}}\Biggr),
\eqno(3)
$$
where the spin kinetic energy operator and the spin-current operator
along the $x$-direction are defined by $T_x^{(s)} = \sum_{\bf
i}{1\over 2}J_{\bf i,i+\hat x}(S_{\bf i}^+S_{\bf i+\hat x}^- + {\rm
h.c.})$ and $j_x^{(s)} = \sum_{\bf i}{i\over2}J_{\bf i,i+\hat
x}(S_{\bf i}^+S_{\bf i+\hat x}^- - {\rm h.c.})$, respectively, and
where $N$ denotes the number of sites.

We can next consider placing a small uniform dynamical twist,
$\theta_x(t) = \theta_x e^{i\omega t}$ on all of the nearest-neighbor
bonds that are oriented along the $x$-direction in the modified
Heisenberg model (1).  Application of the Kubo formula then yields a
variation in the ground state energy per site of $N^{-1}\Delta E_0 =
{1\over 2}\Pi_{xx}^{(s)}\theta_x^2$, where
$$\Pi_{xx}^{(s)} =
N^{-1}\Biggl[\langle -T_x^{(s)}\rangle -
\sum_{\nu\ne 0}\Biggl({|\langle 0|j_x^{(s)}|\nu\rangle|^2
\over{E_{\nu}-E_0-\omega}}- {|\langle \nu|j_x^{(s)}|0\rangle|^2
\over{E_{0}-E_{\nu}-\omega}}\Biggr)\Biggr].
\eqno (4)
$$
Since the spin-current is given by $j_x^{(s)} = \partial
E_0/\partial\theta_x = N\Pi_{xx}^{(s)}\theta_x$, the spin conductivity,
$\sigma_s(\omega) = \Pi_{xx}^{(s)}/i\omega$, is then just
$$\eqalignno{ {\rm Re}\, \sigma_s(\omega) =
&2\pi\Biggl[D_s\delta(\omega)+N^{-1}\sum_{\nu\ne 0}|\langle
0|j_x^{(s)}|\nu\rangle|^2\delta((E_{\nu}-E_0)^2-\omega^2)\Biggr], &
(5a)\cr {\rm Im}\, \sigma_s(\omega) =
&\omega^{-1}\Biggl[{1\over{N}}\langle -T_x^{(s)}\rangle -{2\over
N}\sum_{\nu\ne 0} {{|\langle
0|j_x^{(s)}|\nu\rangle|^2(E_{\nu}-E_0})
\over{(E_{\nu}-E_0)^2-\omega^2}}\Biggr].
& (5b)\cr}
$$
 Integrating Eq. (5a) over all frequencies, and substituting in
expression (3) for the spin rigidity, we obtain the following moment
sum-rule$^3$ for the spin conductivity: $$\int_{-\infty}^{\infty}
d\omega {\rm Re}\, \sigma_s(\omega) = \pi {\langle
-T_x^{(s)}\rangle\over {N}}. \eqno (6)$$ Hence, the fraction of the
moment sum-rule occupied by the static twist-response is simply $$I_0
= {\rho_s\over{\langle -T_x^{(s)}\rangle/N}},\eqno (7)$$ where we
define $\rho_s=2D_s$ to be the spin-stiffness.  Notice that the above
result indicates that the entire magnetic moment is made up only of
excited states in unstiff spin systems with $\rho_s=0$.  Below, the
spin-stiffness, as well as the latter static moment-fraction, is
computed numerically using the Lanczos technique on finite chains and
square-lattices for both the ferromagnet and for frustrated
antiferromagnets.

{\it Ferromagnet.} To check the validity of the method we consider
first the nearest-neighbor ferromagnetic spin-1/2 chain with periodic
boundary conditions.  The ground-state of Hamiltonian (1) has been
obtained by applying the Lanczos technique for $N=8, 10, 12, ..., 20$
sites in the $S_z=0$ subspace, that permits introduction of twists
along the $z-$ axis.  After finite-size extrapolation to the thermodynamic
limit as function of $N^{-1}$, we obtain a value for the
spin-stiffness of $\rho_s/J=0.248\pm 0.005$, which is quite close to
the exact result of $\rho_s/J=s^2=1/4$ for spin $s=1/2$.$^{1}$ Similar
results were obtained in the case of the square lattice.  Also, the
average spin kinetic energy for the nearest-neighbor ferromagnet is
simply the total energy in the $S_z=0$ subspace; i.e, $\langle
T_x^{(s)}\rangle=-Ns^2J$.  Hence, Eq. (7) indicates that $I_0=1$,
which means that the static twist-response saturates the moment
sum-rule (6) in the case of the nearest-neighbor ferromagnet. We have
recovered the latter result numerically to within computer accuracy.
The saturation effect can also be directly understood by the
comparison of expression (3) for the spin-rigidity and expression (6)
for the moment sum-rule, coupled with the observation that the
ferromagnetic state is a null eigenstate of the spin-current operator.
Hence, the spin-response of such a ferromagnet is analogous to the
charge-response of non-interacting electrons, where the Drude weight
saturates the $f$ sum-rule.$^{1,2}$

{\it Frustrated Antiferromagnetic Chain.} Consider next a periodic
spin-1/2 chain with both nearest-neighbor ($J_1$) and
next-nearest-neighbor ($J_2$) antiferromagnetic interactions.  Again,
we have performed exact diagonalizations of Hamiltonian (1) in the
$S_z=0$ subspace for $N=8, 10, 12, ...,$ and $20$ sites.  The
stiffness extracted from these studies are shown in Fig.1.  After
performing a finite-size extrapolation of our results for the
unfrustrated antiferromagnetic chain ($J_2=0$) as a function of
$N^{-1}$, we obtain a value of $\rho_s/J_1=0.270\pm 0.005$ for the
spin-stiffness in the thermodynamic limit, that is slightly greater
than the exact value of $1/4$ (see ref. 1).  The small discrepancy we
obtain with respect to the exact result could be due to the fact that
the existence of {\it algebraic} long-range order in the spin-1/2
Heisenberg chain$^{1,8}$ exaggerates finite size effects.$^9$ Also, we
see from Fig. 1 that while the stiffness rises very slightly upon the
introduction of frustration, $J_2>0$, it drops precipitously to zero
around $J_2/J_1 = 0.43$.  This point is in the vicinity of the well
studied spin-Peierls dimerization transition,$^8$ evidenced by the
absence of spin-rigidity in the chain.  Recent estimates that exploit
conformal invariance find a value for the latter critical frustration
of $J_{2c}\cong 0.24 J_1$,$^{10}$ which is consistent with the
decrease of $J_{2c}$ with increasing lattice size that we observe
(see Fig. 1). In fact, finite-size extrapolation of these results
yields a value of $J_{2c}/J_1=0.33 \pm 0.05$.  We have also computed
the fraction, $I_0$, of the moment sum-rule occupied by the static
twist-response, which is shown in the inset to Fig. 1.  In the case of
the unfrustrated antiferromagnetic chain ($J_2=0$), this value
extrapolates to $I_0=0.915\pm 0.005$ in the thermodynamic limit,
$N^{-1}\rightarrow 0$.  It is intriguing to remark that the latter
value is quite close to the analogous fraction of the $f$-sum-rule
contributed to by the Drude weight in the $t-J$ model chain with one
hole,$^{11}$ which is $0.938$.  In addition, this fraction increases
to a maximum value approaching unity at $J_2/J_1\cong 0.25$ of
$I_0=0.986\pm 0.005$, just before plummeting to zero.

{\it Frustrated Antiferromagnet on the Square Lattice.} We have also
diagonalized Hamiltonian (1) for the case of spin-1/2 on finite square
lattices with nearest-neighbor ($J_1$), next-nearest-neighbor ($J_2$),
and next-next-nearest-neighbor ($J_3$) interactions.  This model has
been widely studied because of its close connection with the $t-J$
model on the square lattice,$^{12}$ and hence because of it's relation
to the phenomenon of high-temperature superconductivity.$^{7,13-15}$
In particular, we have found ground states for $N=8,16,18$ and $20$
site square lattices with periodic boundary conditions$^{16}$ via the
Lanczos technique.  In the case of the nearest-neighbor Heisenberg
antiferromagnet we obtain stiffness values of $\rho_s/J_1=0.185,
0.177$ and $0.174$ for systems sizes of $N=16, 18$ and $20$,
respectively.  Notice the general downward trend with increasing
lattice size.  After finite-size extrapolation to the thermodynamic
limit we find a value of $\rho_s/J_1=0.14\pm 0.01$.  Both this value
and the former upper-bound of $0.174$ for the spin-stiffness lie below
that of $\rho_s/J_1=0.18$ obtained from second-order spin-wave
theory.$^5$

Our results for the spin-stiffness of the frustrated $J_1-J_2$ model
($J_3=0$) are shown in Fig. 2, while those of the frustrated $J_1-J_3$
model ($J_2=0$) are shown in Fig. 3. As intuitively expected, we
observe that the spin-stiffness generally decreases smoothly as
frustration increases.  In particular, the stiffness vanishes near
$J_2/J_1=0.5$ in the case of the $J_1-J_2$ model with 20-sites,
whereas it vanishes near $J_3/J_1=0.35$ in the case of the $J_1-J_3$
model.  The latter parameter values are close to the points where the
corresponding classical model looses its long-range N\' eel order.$^7$
The fraction, $I_0$, of the moment sum-rule exhausted by the static
twist-response for both the $J_1-J_2$ and the $J_1-J_3$ models are
also plotted in the insets of Figs. 2 and 3, respectively.  Fig. 2
shows that the $J_1-J_2$ model on the $4\times 4$ lattice has a
distinct feature near it's classical critical point for values of
frustration ranging from $J_2/J_1 \cong 0.55$ to $J_2/J_1 \cong 0.80$.
Here, the spin-stiffness rises with increasing frustration, but then
finally vanishes.  It has been pointed out in the literature that
uniform chiral correlations peak near this region,$^{14}$ and that the
excited states in this vicinity are spin singlets.$^{15}$
This feature could therefore corresponds to a phase with uniform chiral
spin order,$^{13}$ since the latter state is characterized by spin-0
excitations.  A spin glass phase, however, is not ruled
out.$^{15}$  Note that the absence of this feature on the other
18 and 20 site lattices that we have
studied could be due to their ``tilted''
nature.$^{16}$  For example, collinear order -- which we know must
occur in the thermodynamic limit for large
values $J_2$$^7$ -- is not possible in such
lattices.
Clearly, similar diagonalizations of Hamiltonian (1) on
a $6\times 6$ lattice are necessary in order to resolve this issue.

In summary, we have extended the theory of the determination of
spin-rigidity via twisted boundary conditions$^1$ to the general case
of quantum Heisenberg models.  This method has been applied for the
first time to the exact diagonalization of frustrated spin-1/2
antiferromagnetic chains and square-lattices.  Notably, we find an
upper bound of $0.174$ for the spin-stiffness of the unfrustrated
antiferromagnetic Heisenberg model on the square lattice that agrees
to within a few percent with spin-wave calculations.$^5$ However, the
extrapolated value of $0.14\pm 0.01$ obtained from finite-size scaling
to the thermodynamic limit is considerably smaller.

Discussions with S. Trugman are greatfully acknowledged.  This work
was performed under the auspices of the U.S. Department of Energy.
One of the authors (JPR) was supported in part by National
Science Foundation grant DMR-9322427.

\vfill\eject
\centerline{\bf References}
\vskip 16 pt

\item {1.}  B.S. Shastry and B. Sutherland, Phys. Rev. Lett. {\bf 65},
243 (1990).

\item {2.}  W. Kohn, Phys. Rev. {\bf 133}, A171 (1964).

\item {3.}  L.P. Kadanoff and P.C. Martin, Ann. Phys. {\bf 24}, 419 (1963).

\item {4.}  C. Lanczos, J. Res. Natl. Bur. Stand. {\bf 45}, 255 (1950).

\item {5.}  J. Igarashi, Phys. Rev. B {\bf 46}, 10763 (1992).

\item {6.}  E. Manousakis, Rev. Mod. Phys. {\bf 63}, 1 (1991).

\item {7.} J. Ferrer, Phys. Rev. B {\bf 47}, 8769 (1993); P.
Chandra, P. Coleman and A.I. Larkin, J. Phys. Cond. Matt.
{\bf 2}, 7933 (1990).

\item {8.} I. Affleck, J. Phys. Cond. Matt. {\bf 1}, 3047 (1989).

\item {9.} S. Haas, J. Riera, and E. Dagotto, Phys. Rev. B {\bf 48},
3281 (1993).

\item {10.} K. Okamoto and K. Nomura, Phys. Lett. A {\bf 169}, 433 (1992).

\item {11.} X. Zotos, P. Prelov\v sek and I. Sega, Phys. Rev. B{\bf 42},
8445 (1990).

\item {12.} M. Inui, S. Doniach and M. Gabay, Phys. Rev. B {\bf 38},
6631 (1988).

\item {13.} H.J. Schulz and T.A.L. Ziman, Europhys. Lett. {\bf 18},
355 (1992).

\item {14.} D. Poilblanc, E. Gagliano, S. Bacci and E. Dagotto,
Phys. Rev. {\bf 43}, 10970 (1991).

\item {15.} A. Moreo, E. Dagotto, T. Jolicoeur,
J. Riera, Phys. Rev. B{\bf 42}, 6283 (1990).


\item {16.}  J. Oitmaa and D.D. Betts, Can. J. Phys. {\bf 56}, 897 (1978).
\vfill\eject
\centerline{\bf Figure Captions}
\vskip 20pt
\item {Fig. 1.} Shown is the spin-stiffness, $\rho_s$, of the
frustrated antiferromagnetic chain (in units of $J_1$) as a function of
next-nearest-neighbor frustration, $J_2/J_1$, for various system
sizes.  The inset displays the fraction, $I_0$, of the sum-rule
(6) exhausted by the static twist-response.

\item {Fig. 2.} Above, we display the spin-stiffness of the
square-lattice Heisenberg antiferromagnet (in units of $J_1$)
with only
next-nearest-neighbor (diagonal) frustration, $J_2$, for various
system sizes.  The inset shows the corresponding fraction (7) of the
moment sum-rule exhausted by the static twist-response.

\item {Fig. 3.} Similar to Fig. 2, with the exception that only
next-next-nearest-neighbor frustration, $J_3$, is considered.

\end